# Comment on "Quantum dense key distribution"

Antoni Wôjcik*

Faculty of Physics, Adam Mickiewicz University,

Umultowska 85, 61-614 Poznań, Poland

Abstract

In this Comment we question the security of recently proposed by Degiovanni et al. [Phys. Rev. A 69 (2004) 032310] scheme of quantum dense key distribution.

## I. INTRODUCTION

Recently Degiovanni et al. [1] have proposed (and experimentally implemented) a new protocol of quantum key distribution. As it embeds the advantages of both quantum key distribution scheme and quantum dense coding scheme, the new protocol has been called quantum dense key distribution (QDKD). The novelty of QDKD lies in the fact that it enables legitimate users (Alice and Bob) to exchange two bits – one generated by Alice ($j$) and second generated by Bob ($k$) during a single round of the protocol, i.e. with the use of a single qubit traveling from Alice to Bob and backward. In this way the efficiency (if properly defined) of the protocol could surpass that of e.g. BB84 protocol [2]. The authors analyze the effect of the quantum channel losses and conclude that QDKD is more efficient than BB84 for the quantum transmission probability $p \geq 71\%$. QDKD has been claimed to be secure against all individual attacks. The security condition proposed has the form of an inequality (Eq. (6)) of [1]) involving two parameters – quantum bit error rate $Q$ and probability of correlated results $P_{corr}$. The security proof states that fulfillment of this inequality ensures that mutual information between an eavesdropper (Eve) and legitimate users is less than that between Alice and Bob. In this Comment we question this statement. We present two simple schemes which enables Eve to acquire mutual information exceeding that of Alice and Bob while still fulfilling the above mentioned inequality.

This Comment is organized in the following way. In order to present notation we summarize the QDKD in Sec. II. In Sec. III we present simple counterexample to Eq. (4) of [1]. Secs. IV and V contain our main result i.e. the eavesdropping protocols which show that the security condition given by Eq. (6) of [1] is not sufficient to ensure security of the QDKD protocol. We present an attack on the Alice and Bob key in Sec. IV and V respectively.

## II. QUANTUM DENSE KEY DISTRIBUTION

In this section we summarize the QDKD protocol originally proposed in [1]. Each round of this protocol consists of the following steps:

1. Alice prepares two qubits in modes A and B in a singlet state $|\Psi_{AB}^{-}\rangle$, where $|\Psi_{AB}^{\pm}\rangle = (|0_A\rangle|1_B\rangle \pm |1_A\rangle|0_B\rangle)/\sqrt{2}$. (In the experimental implementation of Degiovanni et al. [1] $|0_K\rangle$ ($|1_K\rangle$) stands for a horizontally (vertically) polarized single photon in a given spatial mode $K$).

2. Alice encodes the value of her key $j$ by performing operation $Z_B^j$ where $Z_B = |0_B\rangle\langle 0_B| - |1_B\rangle\langle 1_B|$ and sends the qubit B to Bob.

3. Bob randomly switches between two modes, namely the message mode (MM) and the control mode (CM).

$4^{MM}$. In MM Bob encodes the value of his key $k$ by performing operation $Z_B^k$ and sends the qubit B back to Alice.

$5^{MM}$. Alice performs measurement which discriminates between the orthogonal states $|\Psi_{AB}^{\pm}\rangle$ and announces publicly its result $m$ ($m=0$ stands for $|\Psi_{AB}^{-}\rangle$, whereas $m=1$ stands for $|\Psi_{AB}^{+}\rangle$).

$6^{MM}$. Alice and Bob calculate the value of their partner's key according to

$$k^{(A)} = m \oplus j$$
$$j^{(B)} = m \oplus k,$$
(1)

where $k^{(A)}$ is the value of Bob's key as seen by Alice and $j^{(B)}$ is the value of Alice's key seen by Bob.

$4^{CM}$. In CM Alice and Bob perform local measurements (projectors on the base $\{|0\rangle,|1\rangle\}$) which, together with the use of classical communication, allows an estimation of the mean value of the parity operator $R = |00_{AB}\rangle\langle 00_{AB}| + |11_{AB}\rangle\langle 11_{AB}|$.

After many rounds of the protocol Alice and Bob estimate quantum bit error rates

$$Q^{(A)} = prob(j \neq j^{(B)})$$
$$Q^{(B)} = prob(k \neq k^{(A)})$$
(2)

and the probability of correlated results $P_{corr} = \langle R \rangle$. Of course in the case of noiseless channel and without eavesdropping $P_{corr} = 0$ and $m = j \oplus k$. It follows that $j^{(B)} = j$, $k^{(A)} = k$ and $Q^{(A)} = Q^{(B)} = 0$. On the other hand, if the values of $Q^{(A)}$, $Q^{(B)}$ or $P_{corr}$ are nonzero then Alice and Bob use these parameters to estimate maximal information available to Eve. The main result of the security proof given in Section III of [1] states that the following inequality (given with the use of Shannon binary entropy $H(x)$)

$$H(Q) + H(P_{corr}) < 1, \tag{3}$$

ensures that mutual information between legitimate users is greater then between any of them and the eavesdropper i.e.

$$\begin{aligned} I_{AB}^{(A)} &> I_{AE}^{(A)} \\ I_{AB}^{(A)} &> I_{BE}^{(A)} \\ I_{AB}^{(B)} &> I_{AE}^{(B)} \\ I_{AB}^{(B)} &> I_{BE}^{(B)} \end{aligned} \tag{4}$$

$I_{XY}^{(Z)}$ stands for mutual information on key $Z$ between users $X$ and $Y$.

## III. SIMPLE SCHEME WITH NO CORRELATION INDUCED

With the general strategy of eavesdropping given by operators $J_{BE}$ and $K_{BE}$ acting on mode $B$ and some auxiliary system $E$ in state $|e_E\rangle$ (see Fig. 1 of [1]) the probability of correlated results is given by $P_{corr} = Tr(\rho R)$, where $\rho = J_{BE}(\rho_{AB} \otimes |e_E\rangle\langle e_E|) J_{BE}^+$, with $\rho_{AB} = (|\Psi_{AB}^+\rangle\langle\Psi_{AB}^+| + |\Psi_{AB}^-\rangle\langle\Psi_{AB}^-|)/2$. (Here and below we assume that both keys are symmetric i.e. 0 and 1 is generated with the same probability.) Degiovanni et al. present the following formula (Eq. 4 in [1]) for the probability of correlated results

$$P_{corr} = \frac{1}{2}\left(1 + \langle \mu_{ABE}^+ | \nu_{ABE}^- \rangle\right), \tag{5}$$

where

$$\begin{aligned} |\mu_{ABE}^+\rangle &= K_{BE} Z_B^0 J_{BE} |\Psi_{AB}^+\rangle \otimes |e_E\rangle \\ |\nu_{ABE}^-\rangle &= K_{BE} Z_B^1 J_{BE} |\Psi_{AB}^-\rangle \otimes |e_E\rangle \end{aligned} \tag{6}$$

Below we will show that Eq. (5) is not generally valid. Let us take as an auxiliary system $E$ single mode and let this mode not contain any photons. Such a state will be denoted by $|e_E\rangle = |vac_E\rangle$. Let both operators $J_{BE}$ and $K_{BE}$ be the SWAP operators

$$J_{BE} = K_{BE} = SWAP_{BE}. \qquad (7)$$

Obviously with such an "eavesdropping" scheme Bob detects no photons and consequently $P_{corr} = 0$. On the other hand,

$$\begin{aligned}|\mu^+_{ABE}\rangle &= |\Psi^+_{AB}\rangle \otimes |e_E\rangle \\ |\nu^-_{ABE}\rangle &= |\Psi^-_{AB}\rangle \otimes |e_E\rangle\end{aligned} \qquad (8)$$

and from Eq. (5) one obtains $P_{corr} = 1/2 \neq 0$. Of course, our scheme is not an eavesdropping one as it does not provide any information to Eve. In the next section we show, however, that it can be easily modified to be an effective eavesdropping tool.

## IV. ATTACK ON ALICE'S KEY

The proposed eavesdropping strategy consists of two modes – the eavesdropping mode and error tuning mode. Each photon traveling form Alice to Bob and backward is attacked by one of these two modes at random. The eavesdropping mode is taken with a probability $p$ ($0 \leq p < 1$) and the error tuning mode with a probability $(1-p)$.

*Error tuning mode*. In this mode (see Fig. 1a) Eve does not obtain any information about the key. However, it enables her to control quantum bit error rates $Q^{(A)}$ and $Q^{(B)}$ and thus to control mutual information between Alice and Bob. There is no need for any auxiliary system in this mode. The unitary operator $J_{BE} = I_B$ is just identity on mode $B$, whereas $K_{BE}$ is given by $K_{BE} = Z_B^q$. $q$ takes the value of zero with a probability $\left(\frac{1}{2} + \varepsilon\right)$ and the value of one with a probability $\left(\frac{1}{2} - \varepsilon\right)$. Parameter $\varepsilon$ is bounded by $0 < \varepsilon \le 1/2$. Obviously the probability of the correlated results vanishes i.e. $P_{corr} = 0$. The value of the Alice's final measurement in this mode is given by $m = j \oplus k \oplus q$. It follows form Eq. (1) that $k^{(A)} = k \oplus q$ and $j^{(B)} = j \oplus q$. Thus quantum bit error rates are

$$Q^{(A)} = Q^{(B)} = \frac{1}{2} - \varepsilon \ . \tag{9}$$

*Eavesdropping mode.* In this mode an auxiliary system E is needed (see Fig. 1b). Let it be again be a single, empty mode ($|e_E\rangle = |vac_E\rangle$). The unitaries are defined as

$$\begin{aligned} J_{BE} &= SWAP_{BE} \\ K_{BE} &= Z_B^q \, SWAP_{BE} \end{aligned}. \tag{10}$$

Now $q$ takes both values (zero and one) with the same probability $1/2$. The probability of the correlated results vanishes again ($P_{corr} = 0$). The value of the Alice's final measurement does not depend on $k$ and is given by $m = j \oplus q$. Thus $k^{(A)} = q$, $j^{(B)} = j \oplus q \oplus k$ and the quantum bit error rates are

$$Q^{(A)} = Q^{(B)} = \frac{1}{2} \ , \tag{11}$$

which means that there is no information flow between Alice and Bob in this mode. On the other hand Eve can correctly calculate the value of Alice's key with the use of the equation $j^{(E)} = m \oplus q$. Thus, Eve has a perfect knowledge of Alice's key during the eavesdropping mode and knows nothing about Alice's key during the error tuning mode. It follows that mutual information between Eve and Alice is given by

$$I_{AE}^{(A)} = p \ . \tag{12}$$

From Eqs. (9) and (11) the total quantum bit error rates can be calculated as

$$Q^{(A)} = Q^{(B)} = \frac{1}{2} - x, \qquad (13)$$

where $x = \varepsilon(1-p)$. It leads to the following formula for mutual information between Alice and Bob $I_{AB}^{(A)} = I_{AB}^{(B)} = 1 - H\left(\frac{1}{2} - x\right)$. With the assumed bounds on $p$ and $\varepsilon$ ($0 \leq p < 1$ and $0 < \varepsilon \leq 1/2$) parameter $x$ can take arbitrary value within the range $0 < x \leq 1/2$. This ensures that mutual information $I_{AB}^{(A)}$ is nonzero i.e.

$$I_{AB}^{(A)} > 0. \qquad (14)$$

For a given $p$ Eve can choose the parameter $\varepsilon$ to allow any value of $I_{AB}^{(A)}$ within the range $0 < I_{AB}^{(A)} < 1 - H(p/2)$. Let us now analyze the security condition given in [1]. Clearly $H(Q) + H(P_{corr}) = 1 - I_{AB}^{(A)}$, which due to Eq. (14) always fulfills the required inequality $H(Q) + H(P_{corr}) < 1$.

It should be noticed that our eavesdropping scheme induces losses which can be observed by Bob during the control mode. Let us denote by $p$ the probability that a photon is successfully transmitted from Alice to Bob in the case of no eavesdropping. The transmission probability as observed by Bob - $p_{obs}$ is suppressed due to Eve's action and takes the value $p_{obs} = p(1-p)$. Thus, the parameter $p_{obs}$ could be used as an eavesdropping witness. Let us emphasize, however, that the security proof based on transmission probability cannot be based on fundamental principles, which is usually expected in the field of quantum cryptography. This is because one cannot be sure that Eve does not replace the original quantum channel by a better one with the transmission probability $p'$ ($p' > p$). In this way Eve can hide induced losses provided that she chooses the value of $p < p_{MAX}$, where $p_{MAX} = (p'-p)/p'$. Thus eavesdropping can be safely performed in the realistic case $p < 1$.

## V. ATTACK ON BOB'S KEY

This attack also consists of error tuning mode and eavesdropping mode (taken with probability $(1-p)$ and $p$ respectively. The error tuning mode is the same as the one presented in Sec. IV. In the eavesdropping mode (see Fig. 1c) Eve uses an auxiliary system $E$, which is a single mode again, however, now it contains a single horizontally polarized photon i.e. $|e_E\rangle = |0_E\rangle$. Eve starts with measuring the qubit in the mode B in the base $\{|0\rangle, |1\rangle\}$. Let us denote the result of this measurement by $t$. Note that after the measurement mode B contains no photons. Next two operations, first $V_{BE}$ and then $X_B^t$, are performed on state $|\Psi_{BE}^{(0)}\rangle = |vac_B\rangle|0_E\rangle$. Unitary operation $V_{BE}$ can be easily implemented with the use of a half-wave plate and a polarizing beam splitter (see Fig. 2). $V_{BE}$ performs the following transformation, when acting on relevant states, ($s = 0, 1$)

$$V_{BE}|vac_B\rangle|s_E\rangle = \frac{-i}{\sqrt{2}}\left(|0_B\rangle|vac_E\rangle + (-1)^s |vac_B\rangle|1_E\rangle\right). \tag{15}$$

The operation $X$ is just the polarization NOT gate ($X = |0\rangle\langle 1| + |1\rangle\langle 0|$). It is performed conditionally when the measured photon appears vertically polarized ($t = 1$). Thus, after performing operation $X_B^t V$ the state of the modes $B$ and $E$ is

$$\left|\Psi_{BE}^{(1)}\right\rangle = X_B^t V \left|\Psi_{BE}^{(0)}\right\rangle = \frac{-i}{\sqrt{2}}\left(|t_B\rangle|vac_E\rangle + |vac_B\rangle|1_E\rangle\right). \tag{16}$$

The above equation explains why the probability of correlated results again vanishes ($P_{corr} = 0$). During the control mode measurement Bob has equal chances of finding no photon in mode $B$ or finding a photon in the appropriate polarization state $|t\rangle$. In the message mode the state $\left|\Psi_{BE}^{(1)}\right\rangle$ is transformed by Bob into

$$\left|\Psi_{BE}^{(2)}\right\rangle = Z_B^k \left|\Psi_{BE}^{(1)}\right\rangle = \frac{-i}{\sqrt{2}}\left((-1)^{kt}\left|t_B\right\rangle\left|vac_E\right\rangle + \left|vac_B\right\rangle\left|1_E\right\rangle\right) \quad . \tag{17}$$

Eve transforms the above state with the use of $\left(X_B^t V\right)^+$ operation into $\left|\Psi_{BE}^{(3)}\right\rangle$. It follows from Eq. (15) that

$$\left|\Psi_{BE}^{(3)}\right\rangle = (-1)^{kt}\left|vac_B\right\rangle\left|(k\,t)_E\right\rangle. \tag{18}$$

Thus the result of the measurement performed by Eve on mode $E$ is given by $n = k\,t$. Finally Eve sends a photon in the state $\left|t_B\right\rangle$ back to Alice. Alice state is now given by

$$\left|(1-t)_A\right\rangle\left|t_B\right\rangle = \frac{1}{\sqrt{2}}\left(\left|\Psi_{AB}^+\right\rangle - (-1)^t\left|\Psi_{AB}^-\right\rangle\right). \tag{19}$$

So Alice's measurement generates results $m$ randomly, independently of $j$ and $k$. Quantum bit error rates during eavesdropping mode are thus equal to $1/2$. It means that analysis of the mutual information between Alice and Bob presented in the previous Section applies to the present case as well. Let us now consider the mutual information between Eve and Bob. In the case of $t = 0$ the result of Eve measurement $n = 0$ (which is obviously independent of $k$). However, in the case of $t = 1$ the result of this measurement $n = k$ gives her perfect knowledge about the key. Both cases are equally probably so the considered mutual information is given by

$$I_{BE}^{(B)} = \frac{p}{2}. \tag{20}$$

Let us recall here that for a given $p$ Eve can arbitrarily tune $I_{AB}^{(B)}$ within the range $0 < I_{AB}^{(B)} < 1 - H(p/2)$. Once again Eve gets the advantage $I_{BE}^{(B)} > I_{AB}^{(B)}$, while fulfilling the security condition $H(Q) + H(P_{corr}) = 1 - I_{AB}^{(B)} < 1$.

Let us also consider the losses induced by an attack on Bob's key. The observed transmission probability is given by $p_{obs} = p(1-p) + p/2$. Thus $p_{obs} \geq p$ if $p \leq 1/2$. In this case there is not only no need for quantum channel improvement but even an appropriate fraction of photons has to be filtered off. In the opposite case $p > 1/2$ Eve can hide induced losses if she uses a quantum channel of transmission probability $p' > p$ and takes $p < p_{MAX}$, where $p_{MAX} = (p'-p)/(p'-1/2)$.

## VI CONCLUSIONS

We have presented two eavesdropping schemes which impair the security of the recently proposed QDKD protocol. Let us emphasize that none of them requires the so-called *unlimited computing power.* In contrary, both schemes can be easily implemented with the use of current technology. If Eve can perform polarization CNOT gate (which is still beyond today's technology) she can use the scheme previously proposed by us [3] in the context of ping-pong protocol [4] to improve her attack on Bob's key. With this scheme, the quantum bit error rate generated by eavesdropping can be reduced from $1/2$ to $1/4$. Let us finally emphasize that we do not criticize the general idea of combining quantum key distribution with quantum dense coding. Note that this idea has been recently successfully exploited by Cai and Li [5] in their quantum cryptographic protocol. What we wish to show in this Comment is that the condition given in Eq. (6) of [1] does not ensure secure transmission of the keys.

## ACKNOWLEDGMENTS

I would like to thank the Polish Committee for Scientific Research for financial support under grant no. 0 T00A 003 23.

*Email address: antwoj@amu.edu.pl

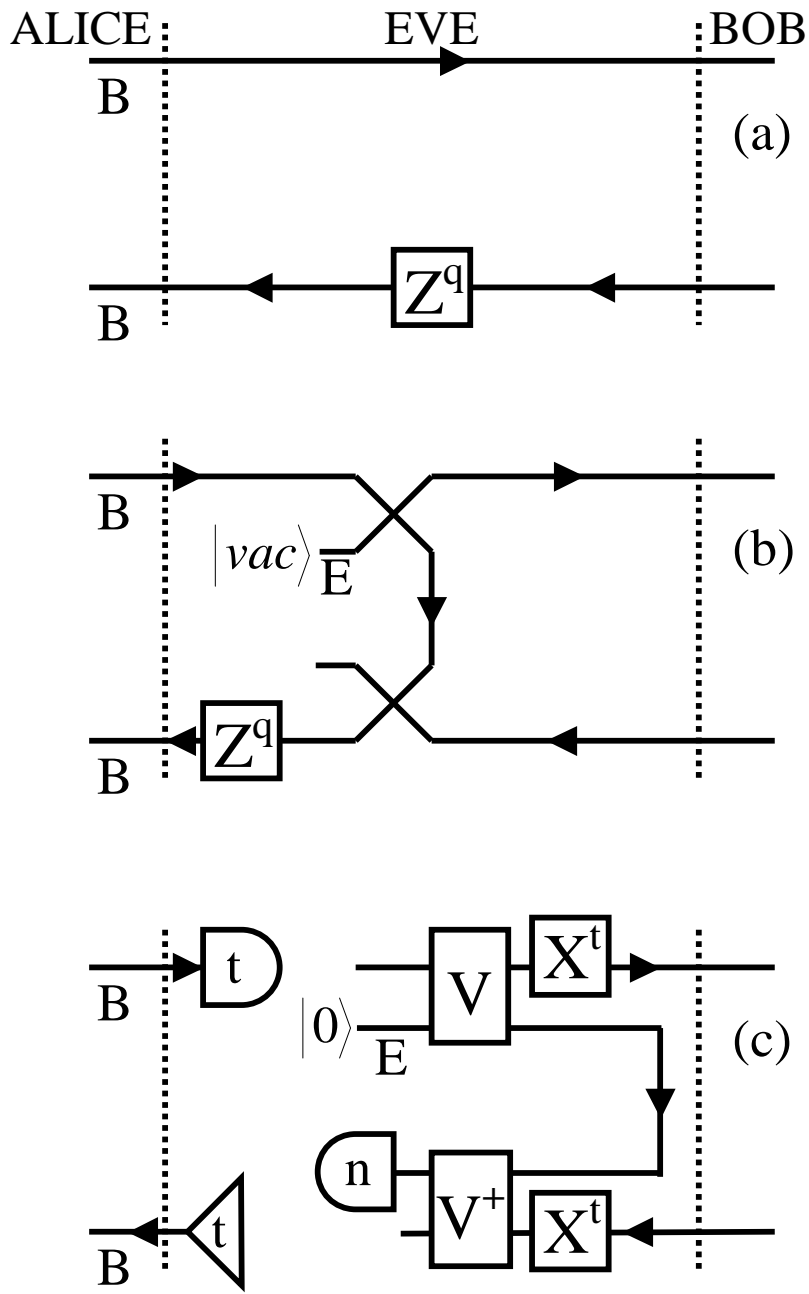

Fig. 1 Eavesdropping schemes: (a) error tuning mode, (b) eavesdropping mode (attack on Alice key), (c) eavesdropping mode (attack on Bob key).

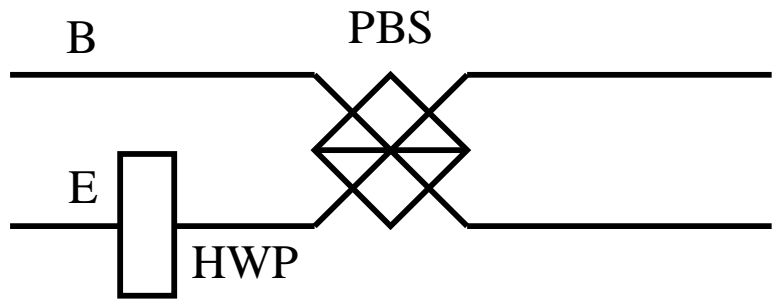

Fig.2 Optical implementation of unitary operation $V_{BE}$. HWP – half-wave plate, PBS – polarizing beam splitter.